\def\wig#1{\mathrel{\hbox{\hbox to 0pt{%
          \lower.5ex\hbox{$\sim$}\hss}\raise.4ex\hbox{$#1$}}}}

\def\v1n{{\cal U}^N_1}

\def\sss{\scriptscriptstyle}
\def\phih2h2{\phi_{\sss {\rm H_2-H_2}}}
\def\phh2{\phi_{\sss {\rm H-H_2}}}

\def\Teff{T_{\rm eff}}
\def\sqr#1#2{{\vcenter{\vbox{\hrule height.#2pt 
  \hbox{\vrule width.#2pt height#1pt \kern#1pt 
  \vrule width.#2pt} 
  \hrule height.#2pt}}}}

\def\wig#1{\mathrel{\hbox{\hbox to 0pt{%
          \lower.5ex\hbox{$\sim$}\hss}\raise.4ex\hbox{$#1$}}}}

\documentclass[preprint2]{aastex}
\input psfig.tex

\shorttitle{Extremely Cool White Dwarfs}
\shortauthors{Oppenheimer et al.}
\begin{document}
\slugcomment{To appear in {\it Astrophysical Journal}}

\title{Observations of Ultracool White Dwarfs}

\author{B. R. Oppenheimer}
\affil{Astronomy Department, University of California,
    Berkeley, CA 94720-3411, USA}
\email{bro@astron.berkeley.edu}

\author{D. Saumon}
\affil{Department of Physics and Astronomy, Vanderbilt University,
    Nashville, TN 37235, USA} 

\author{S. T. Hodgkin}
\affil{Institute for Astronomy, Cambridge University, Madingley Road, 
    Cambridge, CB3 0HA, UK}

\author{R. F. Jameson}
\affil{Department of Physics and Astronomy, University of Leicseter, 
    University Road, Leicester, LE1 7RH, UK}

\author{N. C. Hambly}
\affil{Institute for Astronomy, University of Edinburgh, Blackford
    Hill, Edinburgh, EH9 3HJ, UK}

\author{G. Chabrier}
\affil{C. R. A. L., Ecole Normale Sup\'{e}rieure, 69364 Lyon Cedex 07, France}

\author{A. V. Filippenko, A. L. Coil}
\affil{Astronomy Department, University of California,
    Berkeley, CA 94720-3411, USA}

\and

\author{M. E. Brown}
\affil{Department of Geological and Planetary Sciences, California Institute 
    of Technology, 170-25, Pasadena, CA 91125, USA}

\begin{abstract}
We present new spectroscopic and photometric measurements of the white
dwarfs LHS 3250 and WD 0346$+$246.  Along with F351$-$50, these white
dwarfs are the coolest ones known, all with effective temperatures
below 4000 K.  Their membership in the Galactic halo population is
discussed, and detailed comparisons of all three objects with new
atmosphere models are presented.  The new models consider the effects
of mixed H/He atmospheres and indicate that WD 0346$+$246 and
F351$-$50 have predominantly helium atmospheres with only traces of
hydrogen.  LHS 3250 may be a double degenerate whose average radiative
temperature is between 2000 and 4000 K, but the new models fail to
explain this object.
\end{abstract}


\section{Introduction}

The recent discovery of three white dwarfs with temperatures below
4000 K \citep{ham97,ham99,har99,hod00,iba00} shows that extremely cool
white dwarfs have blue colors in the infrared and may peak in flux
density at wavelengths as short as 0.6 $\mu$m.  The spectral energy
distributions (SEDs) of these cool white dwarfs deviate from the black
body spectrum by up to four magnitudes in the infrared because of the
presence of deep absorption due to collisionally induced dipole
moments in H$_2$ molecules (commonly referred to as H$_2$ collision
induced absorption or H$_2$ CIA; \citet{cia}).  Theoretical
predictions of this phenomenon by \citet{ber93}, \citet{sau99}, and
\citet{han99} have now been unambiguously confirmed in LHS 1126 \citep{ber94},
LHS 3250 \citep{har99} and WD 0346+246 \citep{hod00}, which has been 
spectroscopically observed in the infrared.  The SED demonstrates that
the widespread assumption that optical and infrared colors
monotonically redden with decreasing effective temperature is wrong.

We present here a compilation of new and previously published
observations of three ``ultracool'' white dwarfs, WD 0346$+$246, LHS
3250, and F351$-$50.  Our new observations of WD 0346$+$246 and LHS
3250 include spectra and photometry from $U$ through $K$ band, the
wavelength region in which these stars are brightest.  All of these
objects were discovered because of their high proper motions and all
clearly exhibit extreme flux deficits in the near-infrared.
Furthermore, the measured parallaxes and proper motions enable a
discussion of their membership in the Galaxy's halo population. 

To complement the observational work, we present synthetic spectra of
such white dwarfs.  We expand the models of \citet{sau99} by
considering atmospheres of mixed hydrogen and helium composition.  We
use these new models to fit our new data along with previous parallax
measurements of WD 0346$+$246 \citep{ham99} and LHS 3250
\citep{har99}.  Such fits should ideally provide complete physical 
solutions for these white dwarfs.  However, only the spectra of WD
0346$+$246 and F351$-$50 (published in \citet{iba00} without a
parallax measurement) can be explained by the new models.  LHS 3250
remains an enigma, but we discuss the general properties in depth.
For WD 0346$+$246, because the parallax is known, we derive the
luminosity ($L$), effective temperature ($T_{\rm eff}$), mass 
($M$), and radius ($R$).

\section{Observations}

\begin{figure}[h]
\psfig{file=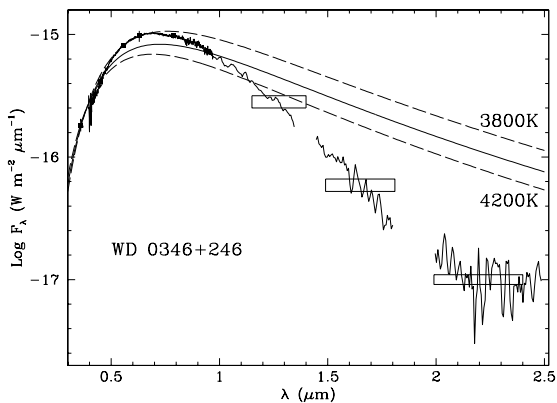,angle=270,width=3in}
\caption{\footnotesize Spectra and photometry of WD 0346$+$246.  The
open boxes indicate the infrared photometry.  Their width corresponds
to the wavelength range of the filter and the height corresponds to
the uncertainty in flux density.  The small closed squares mark the
optical photometric measurements.  Those points without error bars
have errors smaller than the size of the squares.  The spectral data
have been smoothed over 3 pixels (approximately 1 resolution element).
Three black body spectra (at 3800, 4000 and 4200 K) are also plotted
to show the fitting used in \S \ref{sec:physpar} to the wavelength
region below 0.5 $\mu$m.  They also clearly illustrate the strong
infrared suppression.\label{fig:wd0346}}
\end{figure}

\begin{figure}[h]
\psfig{file=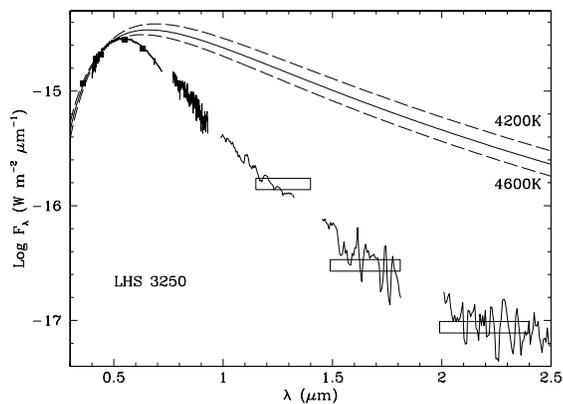,angle=270,width=3in}
\caption{\footnotesize Spectra and photometry of LHS 3250.  The
notation and symbols are identical to those used in Figure
\ref{fig:wd0346}.  The optical photometry and the optical spectrum 
come from \citet{har99}.\label{fig:lhs}}
\end{figure}

Our new observations of LHS 3250 and WD 0346$+$246 spanned more than a
year and involved multiple instruments and telescopes.  For WD
0346$+$246, we have re-reduced our infrared spectrum (first presented
in \citet{hod00}), obtained a new optical spectrum and measured $U$
and $B$ band photometry.  For LHS 3250, we have obtained spectra
spanning 0.8 to 2.5 $\mu$m.  The photometric measurements are shown in
Table \ref{tab:wd0346}.  Figure \ref{fig:wd0346} presents the
photometry and spectroscopy of WD 0346$+$246, and Figure \ref{fig:lhs}
plots the new spectroscopy of LHS 3250 along with the \citet{har99}
spectrum and photometry.  In our subsequent analysis sections we
additionally consider the spectrum of F351$-$50 from \citet{iba00}.

\begin{deluxetable}{crcccc}
\footnotesize
\tablecaption{Photometry of WD 0346$+$246\label{tab:wd0346}}
\tablewidth{0pt}
\tablehead{
\colhead{Band} & \colhead{$\lambda_{\rm eff}$}   & \colhead{Mag.}   &
\colhead{$f_\lambda$} & \colhead{Lum.}\\
\colhead{} & \colhead{($\mu$m)} & \colhead{} & \colhead{(W m$^{-2}$ $\mu$m$^{-1}$)} & \colhead{(L$_\odot$)}
}
\startdata
$U$ & 0.3686 & 20.8 $\pm 0.1 $ & 1.81$\times$10$^{-16}$ & 5.0$\times$10$^{-7}$ \\
$B$ & 0.4310 & 20.5 $\pm 0.1 $ & 4.14$\times$10$^{-16}$ & 1.6$\times$10$^{-6}$ \\
$V$ & 0.5405 & 19.06 $\pm 0.01 $ & 8.51$\times$10$^{-16}$ & 2.5$\times$10$^{-6}$ \\
$R$ & 0.6314 & 18.30 $\pm 0.08 $ & 1.02$\times$10$^{-15}$ & 3.0$\times$10$^{-6}$ \\
$I$ & 0.7927 & 17.54 $\pm 0.02 $ & 1.08$\times$10$^{-15}$ & 4.4$\times$10$^{-6}$ \\
$J$ & 1.2117 & 17.60 $\pm 0.05 $ & 2.79$\times$10$^{-16}$ & 1.7$\times$10$^{-6}$ \\
$H$ & 1.6226 & 18.2 $\pm 0.1 $ & 5.93$\times$10$^{-17}$ & 4.5$\times$10$^{-7}$ \\
$K$ & 2.1750 & 19.0 $\pm 0.1 $ & 9.97$\times$10$^{-18}$ & 1.0$\times$10$^{-7}$ \\
 \enddata

\tablecomments{\footnotesize According to the synthetic spectra (\S \ref{sec:models}),
the total luminosity of this object is about 15\%\ higher than the sum
of the numbers in the last column.}
\end{deluxetable}

\subsection{WD 0346$+$246}\label{sec:wdopt}

\citet{ham97} presented an optical spectrum of WD 0346$+$246
which fell short of reaching 1.0 $\mu$m.  With the infrared spectrum
in \citet{hod00}, a gap in the data set persisted between 0.8 and 1.0
$\mu$m.  The \citet{hod00} infrared spectrum seemed to indicate the
presence of a sharp peak in the spectrum somewhere in this gap.  The
\citet{ham97} spectrum was about 0.1 mag discrepant with their optical
photometry as well.  With this in mind, we revisited the reduction of
the infrared spectrum and obtained a new optical spectrum with careful
flux calibration independent of any imaging photometric observations.

The new reduction of the infrared spectral data previously published in
\citet{hod00} (taken February 1999 at the Keck I 10-m telescope with
the Near Infrared Camera, NIRC; \citet{mat94}) revealed an error in
the spectrophotometric calibration which caused a peak near 1.0
$\mu$m.  The error only affected the spectrum in the $Z_{\rm CIT}$
band and minimally in the $J$ band.  The error was due to an incomplete
removal of the spectral shape of the standard star.  This is a broad
effect and introduced no spurious, fine features.  Upon correction,
the optical and near-infrared spectra agree.  Furthermore, the
infrared spectrum agrees perfectly with the new, more carefully
calibrated, optical spectrum described below.

The new optical spectrum of WD 0346$+$246 was taken with the
Low-Resolution Imaging Spectrometer (LRIS; \citep{oke95}) 
on the Keck II 10-m telescope on 15 December
1999 UT under clear conditions and 0.7$^{\prime\prime}$ seeing.  The
spectrum was obtained with a low-dispersion grating and two different
blocking filters to acquire the full wavelength range from 0.39 to 1.0
$\mu$m.  The 150 l mm$^{-1}$ grating blazed at 0.75 $\mu$m in
conjunction with the 1.0$^{\prime\prime}$ slit provided an average dispersion
of $\sim$ 4.8 \AA\ pixel$^{-1}$.  LRIS has a single Tektronix
2048 $\times$ 2048 pixel CCD, which was binned in the spatial
direction by a factor of 2 yielding 0.43$^{\prime\prime}$
pixel$^{-1}$.

A first set of exposures was obtained with the order blocking filter
GG495, which is transmissive only at wavelengths greater than $\sim 0.48$
$\mu$m.  This order-blocking filter is necessary to prevent the
second-order blue part of the spectrum from overlapping the
first-order red spectrum.  Without the blocking filter, only
wavelengths shorter than about 0.74 $\mu$m are uncontaminated.  With
the blocking filter, wavelengths from 0.48 to 0.96 $\mu$m can be
observed without appreciable contamination.  Two exposures of 295 s
each were taken.  Immediately afterward, without repositioning the
telescope, we took three internal (quartz lamp) flatfield exposures
and one exposure with the Xe-Ar lamps to provide the wavelength
calibration.  The standard spectrophotometric calibrator HD 84937 ($V=
8.3$ mag, F5V, \citet{oke83}) was observed with the same settings to
provide the spectral response of the spectrograph at an airmass of
1.12, within 10\%\ of the airmass for WD 0346$+$246.  The position
angle for the slit was chosen to be the parallactic angle to minimize
the effects of atmospheric dispersion \citep{fil82}.

The second set of data was taken with the same grating and slit
settings, but without the blocking filter.  Again two 295 s exposures
were obtained along with corresponding internal flatfields and arc
lamp exposures.  The same standard star was also observed in this
manner.

In order to be certain that the spectrum near 1.0 $\mu$m was correct
and uncontaminated, we obtained service observations with the same
instrument settings on 24 December 1999 UT but with the RG850 blocking
filter installed.  This transmits wavelengths redward of $\sim$ 0.83 $\mu$m,
with the second-order contamination setting in at $2 \times 0.83 = 1.66
\mu$m.  The weather was clear and the seeing was again
$\sim 0.7^{\prime\prime}$.  One 600 s exposure was taken, along
with internal flatfields and arc lamp exposures.  Because of the
low sensitivity of the CCD in the near-infrared, these data are very
noisy.  

Using IRAF, we conducted standard CCD processing and optimal spectral
extraction, while flux calibration and telluric absorption band
removal were done with our own routines in IDL.  Final adjustments to
the wavelength scale were obtained using the background sky regions.

The adopted reduction technique produces a spectrum which is flux calibrated
independent of the photometric measurements.  In this sense the
spectrum differs from the infrared spectrum described above.  The
infrared spectrum is flux calibrated using the infrared photometry and
the shape of a calibrator G star, which is not a spectrophotometric
calibrator.  In that case the G star is used only to provide the
correct spectral shape, while the absolute calibration is provided by the
infrared photometry.  The spectrum presented in Figure \ref{fig:wd0346}
has a signal-to-noise ratio (S/N) over several hundred through the middle of
the optical region.  The infrared S/N falls from
2000 near 1 $\mu$m to approximately 6 in the $K$ band.

The data taken on 15 December exhibits second-order contamination beyond
0.98 $\mu$m, as shown by the uncontaminated data taken on 24 December.
However, the infrared spectrum from NIRC has a much higher
S/N, and the 24 December LRIS spectrum agrees
perfectly with it, so we simply cut the optical spectrum at 0.97
$\mu$m.  Thus, in Figure \ref{fig:wd0346}, the data redward of 0.97
$\mu$m are from NIRC.

The optical photometry of WD 0346$+$246 presented in \citet{hod00} and
\citet{ham97} was derived from CCD images made at the Jacobus Kapteyn
Telescope.  The $B$-band measurement only provided an upper limit and
there was no $U$-band measurement.  We have subsequently obtained $U$ and
$B$ band CCD images of WD 0346$+$246 with the Palomar 1.5-m Oscar Meyer
telescope and a Tektronix 2048$\times$2048 pixel thinned, AR-coated
CCD along with Johnson $U$ and $B$ filters.  The pixel scale was
0.37$^{\prime\prime}$ pixel$^{-1}$, and the seeing on 10 and 11 August
1999 UT was 1.0$^{\prime\prime}$ under photometric conditions.

On 10 August we obtained five 600 s exposures of the WD 0346$+$246
field.  Each exposure was shifted by 10$^{\prime\prime}$ in one of the
cardinal directions from the previous exposure to minimize the effects
of bad pixels.  The \citet{lan92} standard field SA 112 was
also observed with two 7 s exposures such that stars 250, 223, and 275
were in the field of view.

On 11 August 1999, we observed WD 0346$+$246 again but with the $U$
filter.  Four 1200 s exposures were made with dithering procedures
similar to those described above.  For $U$-band photometric standards we
observed the \citet{lan92} field SA 110 and obtained images of
stars 496, 497, 499, 502, 503, 504, 506, and 507 in two 60 s exposures.

Data were reduced by subtracting a median of five dark frames obtained
on each night and dividing by flatfield exposures made by exposing
the CCD to the illuminated dome.  To obtain accurate photometric
measurements of the white dwarf, we averaged the measurements of all
the standard stars visible in each of the two calibration fields
separately. 

\subsection{LHS 3250}

\citet{har99} presented the optical spectrum of LHS 3250 in the
range 0.4 to 0.7 $\mu$m, as well as optical and infrared photometry
(reproduced in Figures \ref{fig:lhs} and \ref{fig:compare}). To
explore the SED in more detail beyond 0.7~$\mu$m, we obtained optical
and infrared spectroscopy.

We measured the 0.75 to 0.98 $\mu$m spectrum of LHS 3250 through
service observations at the William Herschel Telescope in the Canary
Islands. The low resolution spectrum was obtained using the ISIS
spectrograph with the R158R grating giving a wavelength scale of 2.9
\AA\ pixel$^{-1}$ over 2970 \AA\ on the TEK2 1024 $\times$ 1024
pixel detector. The weather was good during the observations, and the 
seeing was
measured to be 1$^{\prime\prime}$. A 1$^{\prime\prime}$ slit was
employed. Two exposures, one of 1800 s and one of 1500 s, were taken
on the target. These were bracketed by exposures of the
spectrophotometric standard Ross~640. Calibration exposures of a
tungsten flat (for flatfielding) and Cu-Ar and Cu-Ne arc lamps (for
wavelength calibration) were also made. CCD data reduction, optimal
spectral extraction, and flux calibration were performed using IRAF
tasks.

We also observed LHS 3250 on 21 July 1999 UT using the Keck I 10-m
telescope and NIRC \citep{mat94}.  We obtained $Z_{\rm CIT}$, $J$, $H$,
and $K$-band spectra spanning 1.0 to 2.5 $\mu$m.

The spectra were obtained at airmass $\sim 1.4$ during a
photometric night with 0.3$^{\prime\prime}$ seeing in $K$ band.  We used
two different settings.  The first measured the $Z_{\rm CIT}$, $J$, and $H$
bands simultaneously using the 150 l mm$^{-1}$ grism and the $JH$
blocking filter.  The slit was 0.3$^{\prime\prime}$ wide.  We obtained
four exposures of 250 s with the white dwarf placed at two different
positions in the slit.

The $H$ and $K$ band spectra were obtained simultaneously with the second
setting.  We used the same slit, the $HK$ blocking filter and the 120 l
mm$^{-1}$ grism.  We obtained five exposures of 200 s each, again placing
the white dwarf in one of two slit positions for each exposure.

The F8V star SAO 17455 ($V = 9.58$ mag) was observed with the same
settings immediately after LHS 3250 to serve as a standard at an airmass
of 1.4, matched to the LHS 3250 observations within 1\%.  For the
first setting we obtained two 4.1 s exposures and for the second
setting three 4.1 s exposures.  

Data reduction followed our standard procedure. We first subtracted
pairs of images from each other to remove the sky background.  Using
the optimal extraction technique we then reduced the spectra to one
dimension.  These steps were carried out on both the object and
standard star spectra.  After calibrating the wavelength scale, using
the well-defined edges of the filter transmission curves (a technique
good to an accuracy of 0.006 $\mu$m over the band), we summed the LHS
3250 spectra and divided by the standard star spectra.  This removes
the pixel-to-pixel variations and accounts for the instrumental
sensitivity as a function of wavelength.  A model F star spectrum was
then divided into the resultant spectra.

\section{Summary of Observations of Ultracool White Dwarfs}\label{sec:sum}

\begin{figure}[h]
\psfig{file=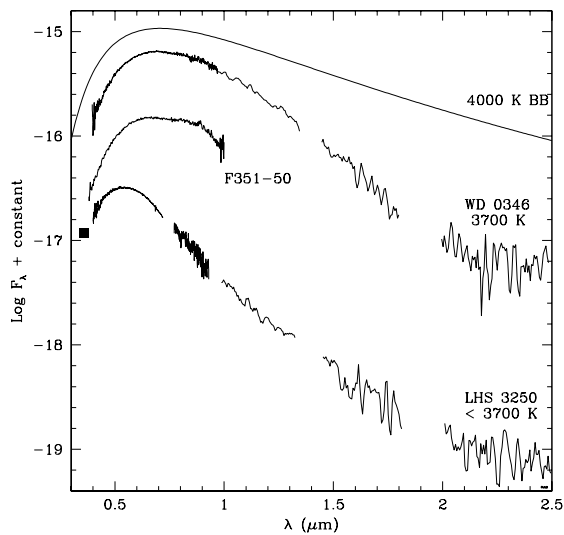,angle=0,width=3in}
\caption{\footnotesize Comparison of spectra of cool halo white
dwarfs; see \S \ref{sec:sum} for a discussion.  The optical spectrum
of LHS 3250 is from \citet{har99} and the F351$-$50 spectrum is from
\citet{iba00}. \label{fig:compare}}
\end{figure}

Figures \ref{fig:wd0346} and \ref{fig:lhs}, and Table \ref{tab:wd0346},
summarize the observations from \S 2.  To provide a means of
comparison, Figure \ref{fig:compare} shows the spectra of LHS 3250, WD
0346$+$246, and F351$-$50 (from \citet{iba00}).  All of the white dwarf
spectra are smooth and exhibit no fine features.  The most important
aspect of this figure is the fact that the three white
dwarf spectra deviate from the black body spectrum redward of 0.6
$\mu$m.  For reference a 4000 K black body spectrum is plotted at the
top of the figure.  The spectra are arranged in order of what we
believe to be decreasing effective temperature, from top to bottom.
(See \S \ref{sec:physpar} for a complete discussion of this.)  Figure
\ref{fig:compare} also shows that the peak wavelength appears to move
toward redder wavelengths, as the temperature drops, until WD
0346$+$246.  Proceeding to cooler temperatures then moves the peak to
{\it bluer} wavelengths.  F351$-$50 appears on this plot because it is
intrinsically faint \citep{iba00} and it exhibits a sudden change in
spectral slope at 0.8 $\mu$m.  According to models (\citet{sau99}; see
\S \ref{sec:models}), this may be the blue edge of the second
vibrational band of the H$_2$ CIA.  Other than this feature, the
spectrum of F351$-$50 is almost identical to that of WD0346$+$246.
The $BVR$ colors are the same for these two objects.  Without infrared
photometry or a parallax it is difficult to make more concrete
statements.  

\section{Are These White Dwarfs Members of the Halo?}

The tangential space motion of WD 0346$+$246 is extremely
high at 170 km s$^{-1}$ and in a direction such that its Galactic
orbit cannot be circular and must be strongly inclined to the Galactic
disk.  The conclusion in \citet{hod00} was that this object is clearly
a member of the Galaxy's halo, regardless of what its unmeasureable
radial velocity is.

\citet{iba00} established that F351$-$50 must have a space velocity
greater than 170 km s$^{-1}$ and is therefore a member of the Galactic
halo.  

For LHS 3250 the case is not nearly as clear.  \citet{har99}
demonstrated that the tangential velocity of LHS 3250 is 81.2 $\pm$
1.2 km s$^{-1}$.  Unfortunately, LHS 3250 is essentially at 90$^\circ$
Galactic longitude and near the plane of the Galaxy.  This, combined
with the lack of any features in the spectrum that would permit
measurement of the radial velocity, makes it impossible to determine
whether it has a disk or halo orbit.  However, there are several
important features to note about the motion of LHS 3250 through the Galaxy.
By subtracting the heliocentric motion from the velocities, it appears
that LHS 3250 is moving toward the south Galactic pole at 73 km
s$^{-1}$ and toward the Galactic anticenter at 52 km s$^{-1}$.  Thus
the only definitive statement is that its orbit about the
Galactic center cannot be circular.  This makes LHS 3250 a probable
member of the halo.

\section{Physical Characteristics}\label{sec:physpar}

In \citet{hod00} we derived the radius, mass, luminosity and effective
temperature of WD 0346$+$246.  This derivation used a technique that
exploits the brightness temperature of these objects as a function of
wavelength.  We have conducted an extensive series of tests of the
technique and have found that it is not reliable.  Here we describe
the technique more explicitly than before, and we also explain why it
is not viable.

\begin{figure}[h]
\psfig{file=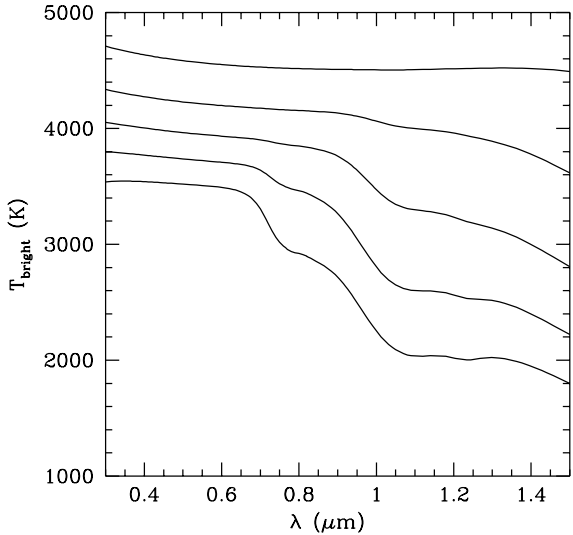,angle=0,width=3in}
\caption{\footnotesize Brightness temperature as a function of
wavelength from pure hydrogen synthetic spectra \citep{sau99}.  All
models have $\log g=8$ and the effective temperature ranges from
4500$\,$K to 2500$\,$K in steps of 500$\,$K, from top to bottom.  In
the cooler models, the strong drop of $T_{\rm bright}$ in the infrared
is due to H$_2$ CIA absorption.  Correspondingly, $T_{\rm bright}$ is
significantly larger than $T_{\rm eff}$ at shorter wavelengths in
these models.  Note the virtual constancy of the brightness
temperature at blue wavelengths. \label{fig:tbright}}
\end{figure}

In \citet{hod00} the derivation of radius, mass, and effective
temperature relied upon the distance, $d$, and the assumption that the
bluest part of the optical spectrum behaves as a black body.  By
fitting this region with a known black body spectrum we believed that
we could obtain the physical parameters.  For pure H white dwarfs, the
only known opacity in this wavelength region is due to H$^{-}$, which
has a weak wavelength dependence in the optical regime.  In pure He
white dwarfs, the dominant opacity source in the optical is He$^{-}$,
which is also very flat.  The flatness of the opacity seemed to
validate the technique.  To support this argument, we examined plots
of the brightness temperature vs. $\lambda$ of synthetic spectra for
pure hydrogen white dwarfs (Saumon \& Jacobson 1999).  Shown in Figure
\ref{fig:tbright}, these plots have nearly constant brightness
temperature in the visible part of the spectrum.  For example, the
brightness temperature of a $\Teff=3500\,$K, $\log g=8$ model, where
$g$ is the surface gravity in cgs units, varies by less than 200$\,$K
from 0.4 to 0.9$\,\mu$m.  For white dwarfs with $\Teff \wig> 4500\,$K,
the SED is quite close to that of a black
body.

To extract the radius of the star, then, involves fitting a black body
curve, $B(T)$, to the optical region where the brightness temperature
is nearly constant (below $\lambda = 0.5 \mu$m).  The fit involves two
parameters, the temperature of the black body fit and the ratio
$(R/d)^2$.  The temperature obtained from the fit is the brightness
temperature in this region of the spectrum, not $T_{\rm eff}$.
Furthermore, the brightness temperature determines the shape of the
black body curve, while the actual flux level is determined by
$(R/d)^2$.  Figures 1 and 2 show attempts to do this fitting on WD
0346$+$246 and LHS 3250.

Using this fitted temperature, $T_{\rm fit}$, in each case, we derive
the ratio $R/d$ from the following equation: $(R/d)^2 \pi
B_{U,B}(T_{\rm fit}) = F_{U,B}$, where $F_{U,B}$ is the flux in the
fitted region from $U$ to 0.5 $\mu$m, and $B_{U,B}(T_{\rm fit})$ is
the black body flux in the same wavelength region.  For LHS 3250, this
yields the result $R = 0.014 R_\odot \pm 0.002$, using the
\citet{har99} value of $d = 30.3 \pm 0.5$ pc.  With the mass-radius
relation from \citet{chab01} we find the mass to be 0.54 $M_\odot$.
For WD 0346$+$246, we derive $R = 0.013 R_\odot \pm 0.002$ and $M =
0.57 M_\odot$.

We can now determine the effective temperature from the relation $L =
4\pi R^2 \sigma T_{\rm eff}^4$.  By integrating all the flux measured
from the $U$ through $K$ bands, we find for LHS 3250 $L = 3.26 \times
10^{-5} L_\odot$ and $T_{\rm eff} = 3650$ K.  Our value for $L$ is
somewhat higher than that derived by \citet{har99}, but it includes
the important region of flux at 1 $\mu$m, for which they had no data.
For WD 0346$+$246, we calculate $L = 1.83 \times 10^{-5} L_\odot$ and
$T_{\rm eff} = 3750$ K.

Criticism about the validity of this method prompted us to test it
thoroughly.  The white dwarf LHS 542 is an excellent test case since
it has been analyzed by \citet{leg98} by model fitting of the
SED, an optical spectrum is available (Ibata
et al.  2000) and the parallax is known.  Our surface brightness
method predicts a radius about 50\% larger than the Leggett et
al. (1998) solution.  We have also performed a variety of tests using
our new models (\S 6) to generate artificial, noiseless spectra
(assuming $R$ and $d$) and comparing the results of the surface
brightness method with the input parameters of the synthetic data.  We
find that the method works very well when the brightness
temperature of the star remains fairly constant over the wavelength
interval of the fit.  When the (monotonic) variation of the brightness
temperature exceeds 100 K, however, the method becomes unreliable even
though satisfactory fits of the spectrum can be obtained (Figures 1
and 2).  The reason is that the derived value of $T_{\rm fit}$ over
the interval is not an ``average'' or intermediate value in the range
of brightness temperatures represented in the fitting interval.  If
$T_{\rm fit}$ were an intermediate value or average of the brightness
temperature, the technique would be viable.  The fact is that $T_{\rm
fit}$ is systematically shifted {\it outside} of the actual range of
brightness temperatures.  This is due to the temperature dependence of
the black body flux just blueward of the peak.  The method thus
converges to anomalous solutions for $T_{\rm fit}$ and $R$, solutions
which deviate increasingly from the actual values as the variation in
brightness temperature increases.  Since the degree of variation of
the brightness temperature in an observed spectrum is {\it a priori}
unknown, it is not possible to apply the method with confidence.

We conclude that the physical parameters of cool white dwarfs can only
be obtained from fitting synthetic spectra to the observed SED.

\begin{figure}[htb]
\psfig{file=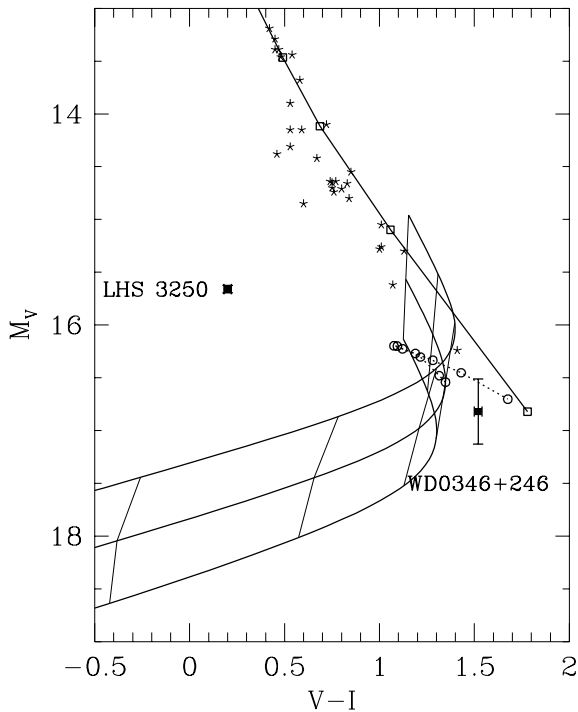,angle=0,width=3in}
\caption{\footnotesize Hertzsprung-Russell diagram for cool white
dwarfs. Solid lines connect the synthetic colors of pure hydrogen
white dwarf atmospheres of a fixed gravity, with $\log g=7.5$, 8, and
8.5 from top to bottom.  Thin lines connect those models which have
the same $\Teff$, starting with $\Teff=4500\,$K right of center and
decreasing in steps of 500$\,$K.  The open circles connected by a
dashed curve show a sequence of atmospheres with mixed hydrogen and
helium composition for $\Teff=3500\,$K and $\log g=8$.  Starting from
the pure H model, the helium-to-hydrogen fraction $\log y =\log N({\rm
He})/N({\rm H})$ increases from $-1$ to 7 in steps of 1.  The colors
of pure He atmospheres with $\log g=8$ are shown by open squares
connected with a solid line.  The value of $\Teff$ increases from 4000
to 7000$\,$K in steps of 1000$\,$K from bottom to top along this
sequence. Star symbols represent cool, single white dwarfs with
hydrogen-rich atmospheres as determined by Bergeron et al.\ (1997).
The error bars on the M$_{\rm V}$ values include the uncertainty in
the parallaxes of the stars.\label{fig:hrdiag}} 
\end{figure}

\section{Analysis of the Photometry}\label{sec:models}

\begin{figure}[htb]
\psfig{file=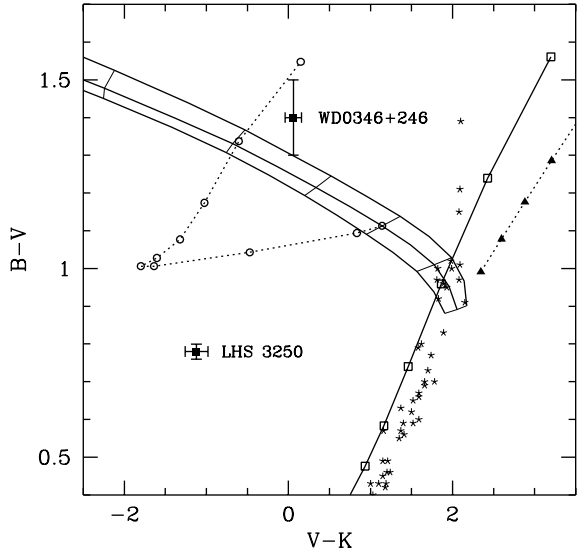,angle=0,width=3in}
\caption{\footnotesize Color-color diagram for cool white
dwarfs. See Figure \ref{fig:hrdiag} for an explanation of the symbols.
The effective temperature along the hydrogen model grid decreases from
4500$\,$K right of center to 2000$\,$K toward the left in steps of
500$\,$K.  The colors of pure He atmospheres with $\log g=8$ are shown
by open squares.  $\Teff$ increases in steps of 500$\,$K starting from
4000$\,$K at the top.  Triangles connected by a dotted line show the
colors of black bodies starting with $T=4500\,$K at the bottom and
decreasing in steps of 250$\,$K.  Compare with Fig. 15 of
\citet{bwb95} or Fig. 9 of Bergeron et al. (1997).\label{fig:color1}}
\end{figure}

\begin{figure}[htb]
\psfig{file=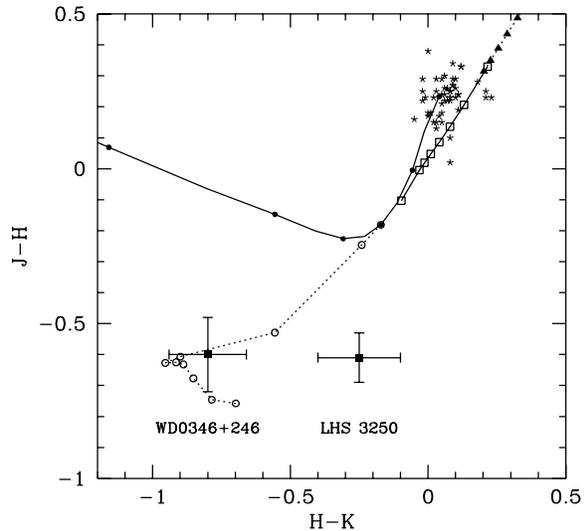,angle=0,width=3in}
\caption{\footnotesize Color-color diagram for cool white dwarfs.
Notation is the same as in Figure \ref{fig:hrdiag}.  Pure H models of
different gravities very nearly overlap in this diagram, so we show
only the $\log g=8$ sequence for clarity.  Filled circles indicate
models with $\Teff=4000\,$K (upper right) and decreasing in steps of
500$\,$K.  Starting from the upper right, the open squares show the
pure He sequence for $\Teff=4000$ to 10000$\,$K in steps of 1000$\,$K,
and $\Teff=20000\,$K at the top-right.\label{fig:color2}}
\end{figure}

We can learn more about these stars by considering their locations in
the Hertzsprung-Russell (HR) diagram (Figure \ref{fig:hrdiag}) and in
color-color diagrams (Figures \ref{fig:color1} and \ref{fig:color2}).
Synthetic spectra for white dwarf atmospheres of pure hydrogen and
pure helium composition provide the means to interpret these diagrams.
In Figures \ref{fig:hrdiag} to \ref{fig:color2}, we show a grid of
pure hydrogen atmospheres (solid lines) with $\log g=7.5$, 8, and 8.5
\citep{sau99}, a sequence of pure helium atmospheres with $\log g=8$
\citep{ber94} (open squares connected by a solid line), and stars from
the cool white dwarf sample of \citet{brl97} which they identified as
having a hydrogen-rich composition and which are not known or
suspected of being binary (star symbols).  Since very cool
white dwarfs cool at constant radius, the sequences of constant
gravity models shown in Figures \ref{fig:hrdiag} to \ref{fig:color2}
also represent cooling tracks.  Along these pure H tracks, we can
associate ages using new cooling models \citep{chab01}.  Table
\ref{tab:ages} gives the ages for the intersection points on the
solid-line grid in Figure \ref{fig:hrdiag}.

Because neither WD 0346$+$246 nor LHS 3250 are explained by the pure
H models, we found it necessary to consider models with mixed H/He
atmospheres (indicated with open circles and dotted lines in Figures
\ref{fig:hrdiag} to \ref{fig:color2}).  These preliminary models have 
fixed values of $T_{\rm eff}$ and $g$ and variable values of the
relative abundance of He ($y = N$(He)$/N$(H)).  These models allow the
exploration of trends resulting from the general effects of mixed
composition.

The qualitative behavior of the mixed H/He atmopsheres can be deduced
from Figure 15 of \citet{ber94}, which shows $B-V$ vs. $V-K$ for
sequences of pure H, pure He, and mixed H and He composition with
$y\le 10$.  As He is added to a cool H atmosphere of a given $\Teff$
and surface gravity, the fractional abundance of H$^-$ decreases, the
relative importance of He-H$_2$ CIA opacity increases, and $V-K$
becomes bluer.  This trend must reverse itself when $y$ becomes very
large, however, since both $V-K$ and $B-V$ of pure He models are very
red (Figure \ref{fig:color1}).  A sequence of models of fixed $\Teff$
and $g$ with composition varying from pure H to pure He is expected to
become first very blue in $V-K$ at roughly constant $B-V$, and then
turn over to redder $B-V$ and $V-K$ until it reaches the colors of a
pure He atmosphere.

Our preliminary sequence of mixed composition atmospheres varies the
value of $y$ over $0.1 \le y \le 10^7$ for a model with fixed
$\Teff=3500\,$K and $\log g=8$.  This sequence is indicated by the
open circles connected by dotted lines in Figures \ref{fig:hrdiag} to
\ref{fig:color2}.  These models are rather crude in their treatment of
the chemical equilibrium and they ignore the non-ideal contributions
to the equation of state which are significant for pure He atmospheres
\citep{ber94}.  Nevertheless, they illustrate the qualitative behavior
of mixed atmospheres and are adequate for the present purpose.  It is
readily apparent that the H$_2$ CIA reaches maximum strength for $\log
y \approx 2$ -- 3.  However the colors of models of very large He
abundance still differ significantly from those of pure He models
because of a residual H$_2$-He CIA opacity and He$^-$ free-free
opacity enhanced by the ionization of H.  We find that in the $BVK$
color-color diagram (Figure \ref{fig:color1}), very cool white dwarfs
with mixed H/He atmospheres occupy a much wider region than, and
overlap with, the pure H atmospheres.  Also, very cool white dwarfs
with $J-H \wig< -0.25$ have atmospheres of mixed H/He composition
(Figure \ref{fig:color2}) and {\it do not} fall between the pure H and
pure He sequences.

\begin{deluxetable}{cccc}
\footnotesize
\tablecaption{Ages of Pure H, Ultracool White Dwarfs\label{tab:ages}}
\tablewidth{0pt}
\tablehead{
\colhead{$T_{\rm eff}$ (K)} & 
\colhead{$\log g = 8.5$} & 
\colhead{$\log g = 8.0$} &
\colhead{$\log g = 7.8$} 
}
\startdata
  4500      &      11.7     &        8.6        &          6.0\\
  4000      &      12.2     &       10.0        &          7.7\\
  3500      &      12.6     &       11.1        &          9.2\\
  3000      &      13.0     &       12.5        &         10.7\\
  2500      &      13.3     &       13.6        &         12.1\\
  2000      &      13.5     &       14.8        &         13.7\\
\enddata
\tablecomments{\footnotesize Ages are given in Gyr and correspond to 
the intersection points in the grid of solid curves in Figure
\ref{fig:hrdiag}.  The last column corresponds to $\log g =7.8$ while 
the bottom curve in Figure \ref{fig:hrdiag} is for $\log g = 7.5$.
All of these ages are the result of new calculations \citep{chab01}
and are only pertinent to pure H atmospheres.}
\end{deluxetable}

Figures \ref{fig:hrdiag} and \ref{fig:color1} show that white dwarfs
with atmospheres of pure He and pure H nearly overlap down to $\Teff
\approx 3500-4000\,$K.  At these low values of $\Teff$, the CIA of H$_2$
becomes very strong and the hydrogen sequence turns over
\citep{han99,sau99}.  The H-rich stars of \citet{brl97} reach down to
the split between the two sequences. In the infrared (Figure
\ref{fig:color2}), however, the two sequences overlap only for very
different ranges of $\Teff$ and are therefore easily distinguished.
The HR diagram (Figure \ref{fig:hrdiag}) is useful in assessing the
evolutionary state of our objects, while the $BVK$ diagram (Figure
\ref{fig:color1}) provides a measure of the overall shape of the
SED.  Because CIA by H$_2$ gives rise
to an opacity that is stronger in the infrared, its effects are
dramatic in $V-K$.

The $JHK$ color-color diagram (Figure \ref{fig:color2}) reveals the
shape of the SED in the infrared and is a most useful diagnostic of
atmospheric composition of cool white dwarfs ($\Teff \wig< 4500\,$K).
In this diagram, both the pure H and pure He sequences are insensitive
to gravity.  This confines atmospheres with pure compositions to two
well-separated curves.  Pure He atmospheres with $\Teff \wig< 4500\,$K
are confined to the upper-right region of the diagram where $H-K >
0.15$ and $J-H > 0.25$, and their colors become redder for lower
$\Teff$.  On the other hand, pure H atmospheres have bluer colors with
$H-K < 0.05$ and $-0.2 < J- H < 0.25$.  The overlap of mixed H/He
atmospheres with the pure composition atmospheres is very limited in
the $JHK$ diagram, and stars with $J-H < -0.25$ most certainly have
mixed composition atmospheres.

Finally, we note that \citet{brl97} discuss the possibility of a
continuum opacity source in pure H atmospheres which would affect the
$U$ and $B$ bands. This helps in reproducing the $B-V$ color of the
most extreme stars in their sample (the uppermost three stars in Figure
\ref{fig:color1}).  We have not included this opacity source in our
models, however.  The spectral fits described below were performed
both with and without the $U$ and $B$ photometry; nearly identical
results were obtained.

It is not presently possible to derive accurate ages for very cool
white dwarfs with pure He or He-rich atmospheres.  However, as a point
of reference we include in Table \ref{tab:ages} ages of extremely
cool, pure H white dwarfs.  Correct treatments of helium pressure
ionization, which determines the location of the photosphere of these
objects, and of He$^-$ absorption cross-sections at high densities,
are lacking and prevent the derivation of reliable cooling curves for
mixed-composition objects.  The only certainty is that, because of
their more transparent atmospheres, white dwarfs with He-rich
atmospheres cool faster than their H-rich counterparts.  For example,
current models \citep{chab97} allow us to estimate that a pure He
atmosphere, 0.6 $M_{\odot}$ white dwarf will reach an effective
temperature $T_{\rm eff} = 4000$ K after only $\sim 6$ Gyr, much
faster than cool DA white dwarfs.  One should further note that even a
small admixture of hydrogen or metals, due either to accretion or
internal mixing processes, will slow down the cooling of the object
dramatically, bringing it closer to a DA cooling sequence.

\subsection{WD 0346+246}

\begin{figure}[htb]
\psfig{file=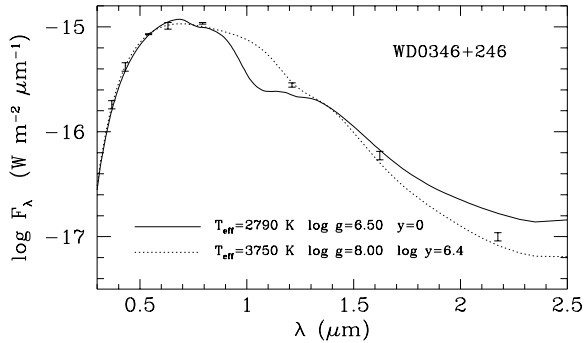,angle=270,width=3in}
\caption{\footnotesize Fit of the $U$ through $K$ photometry of
WD 0346+246.  The broad-band photometric measurements (Table
\ref{tab:wd0346}) are shown by $\pm 1 \sigma$ error bars.  The solid
curve indicates the best-fit pure hydrogen atmosphere model with a gravity of
$\log g=6.5$.  A slightly better fit can be obtained at even lower
gravities.  The dotted line is the best-fit mixed H/He
atmosphere.\label{fig:wdfit}}

\end{figure}

Figure \ref{fig:hrdiag} reveals that WD 0346+246 is intrinsically the
faintest (in the $V$ band) white dwarf known, with the exception of
the high-mass degenerate ESO 439$-$26 \citep{rui95}.  While it is at the
fork between the H and He cooling sequences, it does not obviously
belong to either.  Its $BVK$ colors (Figure \ref{fig:color1}) reveal
that it definitely does not have a pure He atmosphere.  It does not
sit on the pure H models either, except possibly with a very low
gravity of $\log g \approx 6.5$ and $\Teff \approx 2700\,$K.  Indeed, a fit
of pure hydrogen models to the broad-band photometry obtained by
minimizing the error-weighted $\chi^2$ gives $\Teff= 2790\,$K for
$\log g=6.5$ (solid line in Figure
\ref{fig:wdfit}).  Since the distance of WD 0346+246 is known, the fit
directly gives the radius of the star, $R= 0.0174 R_\odot$.  While our
fit is fairly good, it is problematic in several ways.  First, the
optimal $\chi^2$ and $\Teff$ decrease slowly for even lower gravities.
Such low values of $\Teff$ and gravity would be rather extreme
parameters for a white dwarf.  Second, the slope of the fitted
spectrum in the infrared is rather shallow and it misses the $J$ and
$K$ photometry by 3$\sigma$ to 4$\sigma$.  Indeed, the $J-H$ color of WD
0346+246 (Figure \ref{fig:color2}) is $\sim 0.5$ mag bluer than both
the sequences of pure H and pure He models.  Third, the model spectrum
predicts a strong CIA depression near 1$\,\mu$m which is not present
in the observed spectrum (Figure \ref{fig:wd0346}), indicating that
pure hydrogen models are inappropriate.

We think that it is far more likely that the atmosphere of WD 0346+246
is of mixed hydrogen and helium composition.  The colors of WD 0346+246
can be interpreted as those of a mixed H/He atmosphere based on Figures
\ref{fig:hrdiag} through \ref{fig:color2}.  This is particularly 
supported by its $J-H$ color (Figure \ref{fig:color2}).  Using a very
limited grid of mixed composition atmospheres with $\Teff =3500$ and
$3750\,$K, $\log g=8$, and $0 \le y \le 10^7$, we have fitted the He
abundance to the observed broad-band fluxes of WD 0346+246.  The
result is shown by the dotted line in Figure \ref{fig:wdfit}. The fit
of the photometry is excellent and the synthetic spectrum agrees
remarkably well with our observed spectrum (Figure \ref{fig:wd0346})
at all wavelengths.  We obtain a helium abundance of $\log y \approx
6.4$.  Because the models are rather crude, this determination is
inaccurate but it clearly indicates that the atmosphere contains only
traces of hydrogen.  The fitted radius is $R=0.010 \pm 0.002 R_\odot$.
Finally, our choice of $\Teff$ and $\log g$ for the mixed composition
sequence appear to be reasonable estimates for WD 0346+246.  From our
analysis based on model atmospheres and spectra, we conclude that WD
0346+246 is a very cool white dwarf with $\Teff \approx 3700\,$K, a
typical surface gravity of $\log g \approx 8$, and a helium-dominated 
atmosphere ($5 \wig< \log y \wig< 8$).  

\subsection{LHS 3250}

LHS 3250, on the other hand, stands apart from all other known white
dwarfs in Figures \ref{fig:hrdiag} to \ref{fig:color2}.  We can say
with certainty that it does not have a pure He atmosphere.  Its loci
in each of these figures lead to contradictory interpretations.
Figure \ref{fig:hrdiag} suggests that it is a very cool ($\Teff
\approx 2400\,$K), pure H or mixed H/He white dwarf.  Its high 
$V$-band luminosity indicates that it either has a very low gravity and
low $T_{\rm eff}$, or high gravity with very high $T_{\rm eff}$, or it
may be a double degenerate.  We can immediately exclude the high
temperature possibility: its $BVR$ colors are identical to those of WD
1656$-$062, which has a pure H atmosphere, $\Teff=5520\,$K and $\log
g=8$ \citep{brl97}.  If it were this hot, however, it would show
detectable H$\alpha$ absorption but it does not.  Furthermore, its
infrared colors would not be blue.  Its $BVK$ colors (Figure
\ref{fig:color1}) can be understood if $\Teff \approx 4200\,$K with a
mixed H/He atmosphere dominated by He ($\log y \wig> 1$; see also
Fig. 15 of \citet{ber94}).  Its infrared colors (Figure
\ref{fig:color2}) suggest that it has a mixed H/He atmosphere
dominated by He but with a very low temperature.  Finally, our
spectrum is featureless and the absence of a steep drop in the flux
near 1.0 $\mu$m argues against significant H$_2$ CIA, but not against
He-H$_2$ CIA.  It seems that neither pure H nor pure He atmospheres
are sufficient to understand the nature of LHS 3250.  Furthermore,
Harris et al. (1999) were not able to fit its $V$ through $K$
photometry with either pure H or mixed H/He atmospheres.  We have also
failed with our new models and its SED remains unexplained.

A feasible solution to this problem may be that LHS 3250 is an unequal
mass binary, with one component similar to WD 0346$+$246 and the other
much cooler.  This would have the effect of amplifying the optical
flux, as compared to a single white dwarf, while contributing minimal
amplification in the infrared.  Until the full grid of mixed H/He
models has been calculated, a synthetic, composite spectrum of such a
combination is not forthcoming.  We can only conclude that $T_{\rm
eff}$ is below 4000 K, and, based on the severity of the 0.6 to 2.5
$\mu$m absoprtion, we suggest that it must be cooler than WD
0346$+$246 with $T_{\rm eff} \approx 3700$ K.

\begin{figure}[htb]
\psfig{file=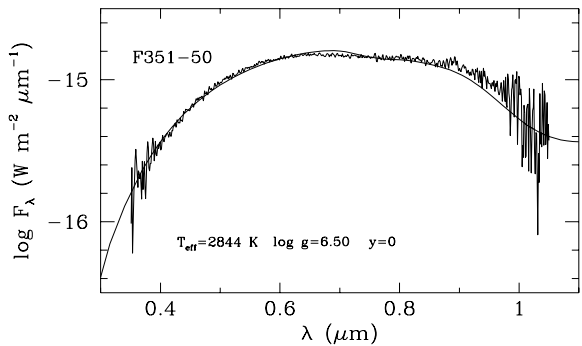,angle=270,width=3in}
\caption{\footnotesize Fit of the spectrum of F351$-$50 by Ibata et
al. (2000) with a pure hydrogen atmosphere.\label{fig:f351fit}}
\end{figure}

\subsection{F351$-$50}

We have analyzed the 0.35 to 1.05$\,\mu$m spectrum of F351$-$50 obtained
by Ibata et al. (2000).  Unfortunately, this part of the spectrum is
not very sensitive to differences in surface composition.  However,
the presence of CIA, and hence of a significant amount of H, is
suggested by the apparent change in the slope of the spectrum for
$\lambda > 0.95\,\mu$m.  Although this is a very noisy part of the
spectrum, we obtain an excellent fit with pure hydrogen models (Figure
\ref{fig:f351fit}).  The fitted parameters are $\Teff=2844\,$K and
$\log g=6.5$.  These are extreme and unlikely parameters for a white
dwarf and we attach only modest significance to the result, for the
same reasons that a similar fit to WD 0346$+$246 was not adequate.

The $BVRI$ colors of F351$-$50 are nearly identical to those of WD
0346+246, suggesting that it may have a similar composition.  If we
ignore the part of the spectrum beyond 0.95$\,\mu$m, where the
S/N declines dramatically, we obtain an equally good
fit with our restricted grid of mixed H/He composition atmospheres
with $\Teff=3500\,$K, $\log g=8$, and $\log y=5.85$.  As in the case
of WD 0346+246, we find that mixed composition models lead to less
extreme physical parameters.  A more definitive analysis will require
infrared photometry or the parallax of F351$-$50.

\section{Conclusion}

WD 0346+246 appears to be the first very cool white dwarf with
an atmosphere of mixed hydrogen and helium composition.  It has an
effective temperature below 4000$\,$K and shows only traces of
hydrogen in its atmosphere. Only three other cool white dwarfs are
known to have significantly high $N({\rm He})/N({\rm H})$ ratios.  All three
are in the C$_2$H spectral class and have $\Teff > 5300\,$K \citep{brl97}.
Although the spectral coverage of the available data is more limited
for F351$-$50, it appears to be very similar to WD 0346+246.  These two
stars are the first examples of very cool white dwarfs with extremely
high He to H mixing ratios and may provide important clues in
understanding the spectral evolution of old white dwarfs.

The computation of an extensive and physically sound grid of mixed
composition atmospheres for $\Teff \wig< 5000\,$K is necessary to
obtain reliable determinations of the surface parameters for
WD 0346$+$246 and F351$-$50. Such a grid of models may provide a better
understanding of the intriguing white dwarf LHS 3250, which 
remains inscrutable.

\acknowledgements

We thank S. R. Kulkarni and J. S. Bloom for their
support of this work, M. Irwin for providing us with his
spectral data on F351$-$50, I. N. Reid for sharing the
spectrum of LHS 3250 and J. Liebert for useful discussions and
suggestions.  We also graciously thank the two referees for a thorough
job and for insisting that we check the technique in \S
\ref{sec:physpar}. Financial assistance for this work was 
provided by NASA through
Hubble Fellowship grant HF-01122.01-99A from the Space Telescope
Science Institute, which is operated by the Association of
Universities for Research in Astronomy, Inc., under NASA contract
NAS5-26555.  BRO
acknowledges FUTDI.  DS and AVF are supported by NSF grants
AST-9731438 and AST-9987438, respectively.  STH acknowledges the
support of the Particle Physics and Astronomy Reasearch Council.
ALC is grateful for a National Science Foundation Graduate Research
Fellowship.  Part of this work was done under the auspice of the
ALLIANCE project \#00193 RL between the United Kingdom and France.
The William Herschel Telescope is operated on the island of La Palma by
the Isaac Newton Group in the Spanish Observatorio del Roque de los
Muchachos of the Instituto de Astrofisica de Canarias.  
The W. M. Keck Observatory is operated as a scientific
partnership among the California Institute of Technology, the
University of California and NASA; it was made possible by the
generous financial support of the W. M. Keck Foundation.

\end{document}